\documentclass[prc,unsortedaddress,superscriptaddress,twocolumn,
showpacs,floatfix]{revtex4}
\usepackage[dvips]{graphicx}

\begin{document}

\title{Asymptotic normalization coefficients for $^{8}$B
$\rightarrow$ $^{7}$Be + $p$
from a study of $^{8}$Li $\rightarrow$ $^{7}$Li + $n$}

\author{L.~Trache}
\affiliation{Cyclotron Institute, Texas A\&M University,
College Station, TX 77843-3366, USA}

\author{A.~Azhari}
\affiliation{Cyclotron Institute, Texas A\&M University,
College Station, TX 77843-3366, USA}

\author{F.~Carstoiu}
\affiliation{Cyclotron Institute, Texas A\&M University,
College Station, TX 77843-3366, USA}
\affiliation{Institute of Physics and Nuclear Engineering
Horia Hulubei, Bucharest, Romania}

\author{H.L.~Clark}
\affiliation{Cyclotron Institute, Texas A\&M University,
College Station, TX 77843-3366, USA}

\author{C.A.~Gagliardi}
\affiliation{Cyclotron Institute, Texas A\&M University,
College Station, TX 77843-3366, USA}

\author{Y.-W.~Lui}
\affiliation{Cyclotron Institute, Texas A\&M University,
College Station, TX 77843-3366, USA}

\author{A.M.~Mukhamedzhanov}
\affiliation{Cyclotron Institute, Texas A\&M University,
College Station, TX 77843-3366, USA}

\author{X.~Tang}
\affiliation{Cyclotron Institute, Texas A\&M University,
College Station, TX 77843-3366, USA}

\author{N.~Timofeyuk}
\affiliation{Department of Physics, University of Surrey,
Guildford, Surrey GU2 7XH, England, UK}

\author{R.E.~Tribble}
\affiliation{Cyclotron Institute, Texas A\&M University,
College Station, TX 77843-3366, USA}

\date{\today}

\begin{abstract}
Asymptotic normalization coefficients (ANCs) for $^{8}$Li $\rightarrow$ $^{7}
$Li $+$ $n$ have been extracted from the neutron transfer reaction $^{13}$C($%
^{7}$Li,$^{8}$Li)$^{12}$C at 63 MeV. These are related to the ANCs in $^{8}$%
B $\rightarrow$ $^{7}$Be + $p$ using charge symmetry. We extract ANCs for $%
^{8}$B that are in very good agreement with those inferred from proton
transfer and breakup experiments. We have also separated the contributions
from the $p_{1/2}$ and $p_{3/2}$ components in the transfer. We find the
astrophysical factor for the $^7$Be($p$,$\gamma$)$^8$B reaction to be $%
S_{17}(0)=17.6\pm 1.7$ eV\,b. This is the first time that the rate of a
direct capture reaction of astrophysical interest has been determined
through a measurement of the ANCs in the mirror system.
\end{abstract}



\maketitle 

Recently the SuperK \cite{superk} and SNO \cite{sno} collaborations
have reported measurements of the solar neutrino flux that provide strong
evidence for neutrino oscillations. Both experiments are primarily
sensitive to high energy solar neutrinos from the $\beta$ decay of $^{8}$B,
produced in the $^{7}$Be($p$,$\gamma $)$^{8}$B reaction. Consequently its
reaction rate at solar energies has been the subject of many recent
studies using both direct \cite{hama,strieder,jung,haas} and indirect
techniques \cite{iwasa,azhari01,davids,trache01,schum03}.

Previously, we used ($^{7}$Be,$^{8}$B) proton transfer reactions to measure
the asymptotic normalization coefficients (ANCs) for the $^{8}$B $\rightarrow
$ $^{7}$Be + $p$ process, from which we determined the astrophysical factor $%
S_{17}(0)$ \cite{azhari01}. However, in those measurements, the separate
contributions of the $p_{1/2}$ and $p_{3/2}$ orbitals could not be inferred
from the ($^{7}$Be,$^{8}$B) angular distributions. Thus, we used microscopic
calculations \cite{mukh90} to fix their relative strengths.

Here we report a study of the mirror neutron transfer reaction, ($^{7}$Li,$%
^{8}$Li), at an energy similar to those used in the proton transfer
reactions. $^{8}$B and $^{8}$Li are mirror nuclei, and charge symmetry
implies that the spectroscopic amplitudes for the proton single particle
orbitals entering the $^{8}$B wave function are nearly the same as those of
the neutron single particle orbitals in $^{8}$Li. Indeed, this has been verified by
many theoretical calculations for $^{8}$Li and $^{8}$B using a variety of 
potential models.
Calculations have been 
done using multi-particle shell models 
\cite{khan,brown,kim,benn99,horoi}, microscopic cluster models \cite{varga,csoto,baye},
or a 
three-body cluster model with long-range correlations \cite{grigorenko} 
with different effective interactions. The
absolute values that they predict for the spectroscopic amplitudes
differ.
However, all calculations agree that spectroscopic factors
for the two nuclei are very similar, with differences being smaller than
2-3\%. Moreover it was shown in Ref.\@ \cite{natasha98} that microscopic
calculations of ANCs for these mirror nuclei are very sensitive to the adopted $NN$
potentials, but their ratio is very stable.  

Previously we have shown \cite{mukh01} that the
$^8$B 
overlap function calculated in a single-particle
approach is an excellent approximation to that obtained from microscopic calculations.
Indeed we have used this fact to obtain ANCs for $^{8}$B $\rightarrow$ $^{7}$Be +
$p$ from transfer reactions \cite{azhari01}.
In this single-particle approach, the spectroscopic factor is related to
the ANC by $C^2=Sb^2$ \cite{mukh97} where $b$ is the single-particle ANC.  Thus
the mirror symmetry between the spectroscopic factors,
coupled with the single-particle approximation,
leads to a proportionality between the asymptotic normalization coefficients
in $^{8}$B $\rightarrow$ $^{7}$Be + $p$ and $^{8}$Li $\rightarrow$ $^{7}$Li
+ $n$ (see Eq.\@ (\ref{CtoC})).  

Mirror symmetry has been used frequently to obtain spectroscopic information
pertinent to astrophysics \cite{barker95,kajino89,timo03}, but its
application to direct capture reactions requires care.
Although charge-symmetry breaking effects on the spectroscopic 
amplitudes only arise at the
few percent level, this does not provide any 
relationship between the $^7$Be($p$,$\gamma$)$^8$B proton capture rate
and its mirror reaction $^7$Li($n$,$\gamma$)$^8$Li.  These reactions
proceed via $s$-wave capture at low energies.  Proton captures on $^7$Be
occur only at large separation distances due to the Coulomb barrier,
so their rate at astrophysical energies
can be calculated from knowledge of the amplitude
of the tail of the $^8$B two-body overlap function in the $^7$Be + $p$
channel, \textit{i.e.}\@ the ANC.  In contrast, the absence of any Coulomb
barrier coupled with the dominant $s$-wave capture in the $^7$Li + $n$ system implies
that the amplitude for the mirror neutron capture reaction may have a
substantial contribution from the nuclear interior, and it can not be
calculated from the ANC alone.  Thus, the proportionality between
the ANCs in
$^{8}$B $\rightarrow$ $^{7}$Be + $p$ and $^{8}$Li $\rightarrow$ $^{7}$Li
+ $n$  does not carry over to the direct capture rates.

We have used the neutron transfer reaction $^{13}$C($^{7}$Li,$^{8}$Li)%
$^{12}$C to obtain the ANCs for $^{8}$Li $\rightarrow$ $^{7}$Li + $n$. The
use of a stable beam in this experiment allows the measurement of the
angular distribution with sufficient resolution that we are able to
determine the strengths of the $p_{3/2}$ and $p_{1/2}$ components
separately. Invoking mirror symmetry, we infer the ANCs for $^{8}$B $%
\rightarrow$ $^{7}$Be + $p$ and use them to determine the astrophysical
factor $S_{17}$. This is a new variation of the ANC approach that will also
be useful in other nuclear systems.

The $^{13}$C($^7$Li,$^8$Li)$^{12}$C neutron transfer reaction at 9 MeV/u is
dominated by a direct one-step process in which the last neutron in the
target is picked up by the projectile. The process can be well described in
DWBA \cite{satchler} and, as we show below, the transfer is peripheral at
this energy. In previous publications~\cite{mukh97}, we have given a general
expression for peripheral reactions relating the angular
distribution to DWBA cross sections and the appropriate ANCs. We chose 
$^{13}$C as a target because it has a relatively loosely bound  
neutron in a $1p_{1/2}$ orbital around a tightly bound core and
the $^{13}$C $\rightarrow$ $^{12}$C + $n$ ANC is known.
The differential cross section for the $^{13}$C($^7$Li,$^8$Li)$%
^{12}$C neutron transfer reaction can be written as 
\begin{eqnarray}
\frac{d\sigma }{d\Omega } = S_{p_{1/2}}(^{13}C)\left[ S_{p_{3/2}}(^{8}Li){%
\sigma}_{\frac{1}{2},\frac{3}{2}}^{DW} +S_{p_{1/2}}(^{8}Li){\sigma }_{\frac{1%
}{2}, \frac{1}{2}}^{DW}\right]  \nonumber \\
= \frac{\ (C_{{}^{12}C, {\frac{1}{2}}}^{{}^{13}C})^{2}}{b_{{}^{12}C, {\frac{1%
}{2}}}^{2}}\left[ \frac{(C_{{}^{7}Li, {\frac{3}{2}}}^{{}^{8}Li})^{2}}{%
b_{{}^{7}Li, {\frac{3}{2}}}^{2}}{\sigma}_{\frac{1}{2}, \frac{3}{2}}^{DW}+%
\frac{(C_{{}^{7}Li, {\frac{1}{2}}}^{{}^{8}Li})^{2}}{b_{{}^{7}Li, {\frac{1}{2}%
}}^{2}}{\sigma }_{\frac{1}{2}, \frac{1}{2}}^{DW}\right] ,  \label{dwcs1}
\end{eqnarray}
where ${\sigma }_{\frac{1}{2},\frac{3}{2}}^{DW}$ and ${\sigma }_{\frac{1}{2},%
\frac{1}{2}}^{DW}$ are the DWBA cross sections for the $p_{1/2}$ $\to$ $p_{3/2}$
and $p_{1/2}$ $\to$ $p_{1/2}$ transitions. $S_{j}(X)$ are the spectroscopic
factors in nucleus $X$, $C_{{}Y,j}^{{}X}$ are the ANCs for $X$ $\to$ $Y$ 
+ $n$, and $b{_{{}Y,j}}$ 
are the ANCs of the normalized single particle bound state neutron wave
functions that are assumed in the DWBA calculations.
For a neutron bound to
the core, the Whittaker function appearing in the
asymptotic behavior of the radial wave function in the proton case \cite
{mukh97} must be replaced by the corresponding Hankel function. In the present 
case the calculated angular distributions for the two $j$   orbitals differ 
at small angles, which permits their contributions to be disentangled. 
To determine the ANCs for $^{8}$Li $%
\rightarrow$ $^{7}$Li + $n$, $(C_{{}^{7}Li,{\frac{3}{2}}}^{{}^{8}Li})^{2}$
and $(C_{{}^{7}Li,{\frac{1}{2}}}^{{}^{8}Li})^{2}$ (denoted below as $%
C_{p_{3/2}}^{2}$ and $C_{p_{1/2}}^{2}$), we need to know the ANC $%
(C_{{}^{12}C,{\frac{1}{2}}}^{{}^{13}C})^{2}$.  However, the ratio
of the ANCs in $^{8}$Li can be obtained without using
$(C_{{}^{12}C,{\frac{1}{2}}}^{{}^{13}C})^{2}$.

Charge symmetry
implies that, to a good approximation,
the spectroscopic amplitudes of $^8$Li and $^8$B are the same,
as demonstrated by the theoretical calculations discussed above. 
Consequently, from the relationship
$(C_{{}Y,j}^{{}X})^{2}=S_{j}(X)(b_{{}Y,j}^{{}X})^{2}$ \cite{mukh97},
one can relate the ANCs in $^{8}$B to those in $^{8}$Li, 
\begin{equation}
C_{p_{j}}^{2}(^{8}B)=C_{p_{j}}^{2}(^{8}Li)~{b_{p_{j}}^{2}(^{8}B)}/{%
b_{p_{j}}^{2}(^{8}Li)}.  \label{CtoC}
\end{equation}
The single particle ANCs
differ due to the
different binding energies and the effect of the Coulomb interaction
on the $^8$B radial wave functions.

\begin{figure}[tbp]
\includegraphics*[width=8.3cm]{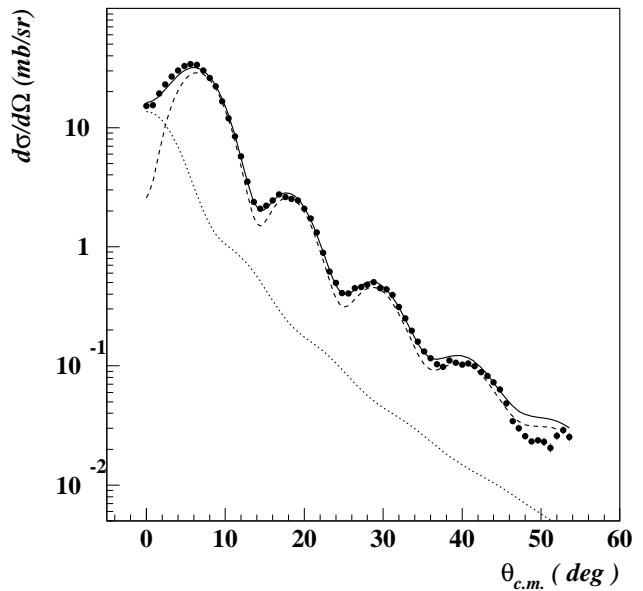}
\caption{The angular distribution for the
$^{13}$C($^{7}$Li,$^{8}$Li)$^{12}$C reaction.
The data are shown as points, and the solid line is the best fit.
The $p_{1/2}$ $\rightarrow$ $p_{1/2}$ component is shown as a dotted line,
and the $p_{1/2}$ $\rightarrow$ $p_{3/2}$ component is the dashed line.}
\label{8Li_angdis}
\end{figure}

The experiment was carried out with a 9 MeV/u beam of $^{7}$Li$^{+1}$ ions
from the K500 superconducting cyclotron at Texas A\&M University. The beam
was transported through the beam analysis system to the scattering chamber
of the MDM magnetic spectrometer, where it interacted with a 300 $\mu $g/cm$%
^{2}$ $^{13}$C target. The target thickness was determined off-line using
the energy loss of $^{228}$Th and $^{241}$Am $\alpha$ sources and
confirmed on-line using the energy loss of the beam. The experimental setup,
including the focal plane detector, was identical to that described in Ref.\@ 
\cite{trache00}. The acceptance of the MDM spectrometer was limited to 4$%
^{\circ}$ in the horizontal by 1$^{\circ}$ in the vertical. An energy
resolution of 120 keV and an angular resolution of 0.18$^{\circ }$, both
FWHM, were obtained for the $^{8}$Li reaction products. Data for the
transfer reaction were obtained for spectrometer settings between $-2^{\circ
}$ and $32^{\circ }$, which covers $0^{\circ }$ to $54^{\circ }$ in the
center-of-mass frame. The angular range $\Delta \theta _{lab}=4^{\circ }$
covered by the entrance slit was divided into eight bins in the analysis,
each point integrating over $\delta \theta _{lab}=0.5^{\circ }$. Typically
we moved the spectrometer by 3$^{\circ }$ at a time, allowing for an overlap
that provided a self-consistency check of the data. The beam current was
integrated with a calibrated Faraday cup at angles larger than 4$^{\circ }$.
For angles around 0$^{\circ }$, we moved the spectrometer in 2$^{\circ }$
steps, and the data were normalized by matching with an overlapping angular
region. This bootstrap approach was used for spectrometer settings out to 4$%
^{\circ }$. Measurements with the spectrometer on both sides of $0^{\circ }$
were made to check beam alignment. The angular distribution for the
population of the $^{8}$Li ground state is shown in Fig.\@ \ref{8Li_angdis}.

\begin{table}[tbp]
\caption{The different optical model parameters used for the DWBA
calculations. The entrance/exit channel parameters were obtained from
phenomenological fits to $^{7}$Li+$^{13}$C, $^{7}$Li+$^{12}$C, and $^{6}$Li+$%
^{13}$C elastic scattering angular distributions, and from the
double-folding procedure. See text for further explanations.}
\label{param}
\begin{tabular}{ccccccc}
\hline
Potential & $V$ & $W$ & $r_{V}$ & $r_{W}$ & $a_{V}$ & $a_{W}$ \\ 
& [MeV] & [MeV] & [fm] & [fm] & [fm] & [fm] \\ \hline
POT1 & 54.3 & 29.9 & 0.92 & 1.03 & 0.79 & 0.69 \\ 
POT2 & 99.8 & 22.0 & 1.01 & 0.77 & 0.81 & 0.81 \\ 
average & 0.366 & 1.00 &  &  &  &  \\ 
fit & 0.323 & 1.00 &  &  &  &  \\ 
$^{7}$Li+$^{12}$C & 97.8 & 18.8 & 0.79 & 0.97 & 0.71 & 0.95 \\ 
$^{6}$Li+$^{13}$C & 77.5 & 16.8 & 0.88 & 1.10 & 0.74 & 0.81 \\ 
JLM-WS & 58.8 & 21.4 & 0.91 & 1.14 & 0.72 & 0.70 \\ \hline
\end{tabular}
\end{table}

DWBA calculations for the transfer reaction were carried out with the code
PTOLEMY \cite{PTOLEMY}. Entrance channel optical model parameters were
obtained by fitting
$^7$Li + $^{13}$C elastic scattering data at 9
MeV/u with a Woods-Saxon form, as reported in Ref.\@ \cite{trache00}. The
potentials labeled 1 and 2 from Table II of Ref.\@ \cite{trache00} were used.
Calculations were carried out using the same parameters for the exit
channel, $^{8}$Li+$^{12}$C. In addition, calculations were done with
entrance/exit channel optical potentials which were obtained from
folding-model potentials using the JLM(1) effective interaction \cite{JLM1}
following the prescription developed in Ref.\@ \cite{trache00}, and with
phenomenological potentials from elastic scattering experiments for similar
systems. A summary of the potentials used is presented in Table \ref{param}.
Parameters from Ref.\@ \cite{trache00} are given in rows 1 through 4. In rows
3 and 4 the renormalization coefficients $N_{V}$ and $N_{W}$ of the folded
potentials are given instead of the potential depth. We used both the
average renormalizations (`average') and those specifically fitted for the $%
^{7}$Li+$^{13}$C case at 63 MeV (`fit'). In rows 5 and 6 we list potential
parameters extracted from neighboring systems
at the same energy per nucleon. The last row
(labeled JLM-WS) was obtained by fitting the exit channel folded potentials
in the surface region ($r=3-12$ fm) with Woods-Saxon shapes and
renormalizing the depths with the average $N_{V}$ and $N_{W}$.

Two components, $p_{1/2}$ $\rightarrow$ $p_{3/2}$ and $p_{1/2}$ $\rightarrow$
$p_{1/2}$, contribute to the $^{13}$C($^{7}$Li,$^{8}$Li)$^{12}$C reaction.
Results of the DWBA calculations using the POT1 entrance and exit channel
potential are shown in Fig.\@ \ref{8Li_angdis}.
The angular distribution for the $p_{1/2}$ $\rightarrow$ $p_{1/2}$
component has a characteristic $l_{tr}=0+1$ shape, while that for the $%
p_{1/2}$ $\rightarrow$ $p_{3/2}$ component has a different $l_{tr}=1+2$
shape. The data obtained for center-of-mass angles between $0^{\circ }$ and $%
30^{\circ}$ allow for a clear separation of the two components. Larger
angles were not used due to increased contributions from multi-step
processes. Combining the two components leads to the solid line fit.

\begin{figure}[tbp]
\includegraphics*[width=8.3cm]{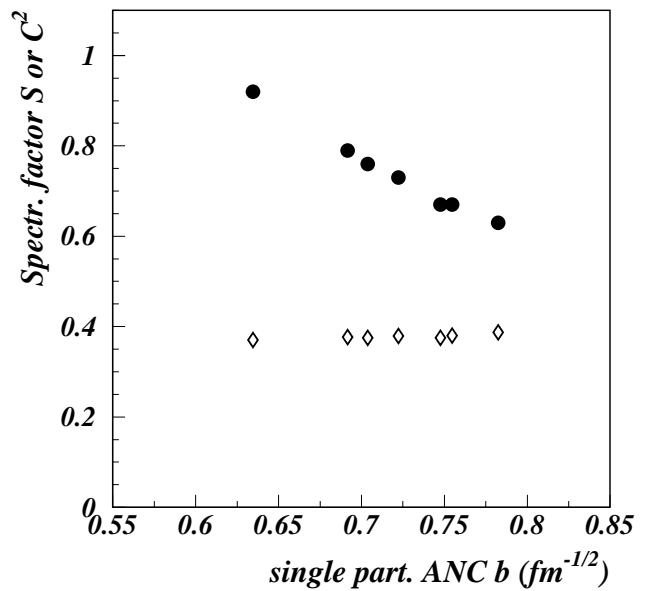}
\caption{Comparison of the spectroscopic factors (dots) and of the ANC $C^2$
(diamonds) extracted in the present experiment, for different geometries of
the single particle Woods-Saxon well. Only the results for the $p_{3/2}$
component are shown.}
\label{8Li_stab}
\end{figure}

In order to verify that the transfer reaction is peripheral,
calculations with the POT1 potential parameters were carried
out using seven different geometries for the
Woods-Saxon potential well that binds the last neutron to the $^{7}$Li core.
Both spectroscopic factors and ANCs were extracted for each calculation.
Figure \ref{8Li_stab} shows the results, plotted against the single particle
ANC $b_{sp}$, for the dominant $p_{3/2}$ component. The spectroscopic
factors vary $\pm 20\%$ around the average, whereas the ANCs vary less than $%
\pm 2\%$, demonstrating that only the asymptotic part of the wave function
contributes in the DWBA calculations and the transfer is peripheral. A
similar result is found for the $p_{1/2}$ component. The ANCs extracted are
therefore independent of the geometry of the single particle potential well
used, whereas the spectroscopic factors are not.

\begin{table}[tbp]
\caption{The results of the present study for different optical model
parameters used for the DWBA calculations. The entrance/exit channel
combinations refer to the potentials in Table \ref{param}. See text for
further explanations.}
\label{results}
\begin{tabular}{cccccc}
\hline
Potentials & $C_{p_{3/2}}^{2}$ & $C_{p_{1/2}}^{2}$ & $\frac{C_{p_{1/2}}^{2}}{%
C_{p_{3/2}}^{2}}$ & $\chi ^{2}$ & angular \\ 
entrance/exit & [fm$^{-1}$] & [fm$^{-1}$] &  &  & fit range \\ \hline
POT1/POT1 & 0.378 & 0.044 & 0.117 & 1.9 & 0-30 deg \\ 
POT2/POT2 & 0.367 & 0.045 & 0.124 & 5.1 & 0-30 deg \\ 
POT1/aver & 0.369 & 0.052 & 0.140 & 5.7 & 0-25 deg \\ 
POT1/aver & 0.379 & 0.052 & 0.139 & 4.8 & 0-20 deg \\ 
aver/aver & 0.363 & 0.049 & 0.136 & 17.4 & 0-30 deg \\ 
aver/aver & 0.384 & 0.054 & 0.140 & 5.7 & 0-20 deg \\ 
fit/aver & 0.390 & 0.053 & 0.136 & 4.6 & 0-20 deg \\ 
fit/fit & 0.376 & 0.053 & 0.141 & 5.8 & 0-20 deg \\ 
POT1/$^{7}$Li+$^{12}$C & 0.370 & 0.044 & 0.118 & 2.5 & 0-30 deg \\ 
POT1/$^{6}$Li+$^{13}$C & 0.409 & 0.047 & 0.115 & 2.9 & 0-30 deg \\ 
POT1/JLM-WS & 0.408 & 0.047 & 0.114 & 3.0 & 0-30 deg \\ 
w. average & 0.384 & 0.048 & 0.125 &  &  \\ \hline
\end{tabular}
\end{table}

Results obtained with different combinations of entrance/exit
channel optical potentials are given in Table \ref{results}. Calculations
done with folded potentials used the JLM(1) potentials with the
corresponding projectile-target combination at the appropriate energy for
each channel and the renormalization values given in Table \ref{param}. The
extracted ANCs are given along with their ratio.
We find $C_{p_{1/2}}^{2}/C_{p_{3/2}}^{2}$ = 0.13(2).
The uncertainty is derived from the
standard deviation of the values obtained for different optical potentials
and from the uncertainties arising from the angular range
used in the fits. This ratio does not depend on the ANC for the ground state
of $^{13}$C or on the absolute values of the individual ANCs in $^{8}$Li,
and is measured for the first time here.

To determine the absolute values of the ANCs in $^{8}$Li, the ANC in $^{13}$%
C was taken to be $(C_{{}^{12}C,{\frac{1}{2}}}^{{}^{13}C})^{2}=2.35\pm
0.12 $ fm$^{-1}$, as calculated from the value of the nuclear vertex
constant, $G^{2}=0.39\pm 0.02$ fm, reported in \cite{natasha}. The results
given in Table \ref{results} show small differences which arise, in
part, from neglecting a small core-core correction in the nuclear part of
the transition operator for the numerical potentials. Differences also arise
from the different renormalizations used, from the inability of the
Woods-Saxon shapes to reproduce the actual shape of double-folded potentials
and from the angular range used in the fits. In particular, the fits with
angular distributions calculated using numerical potentials are not good at
larger angles and consequently have larger $\chi^2$ values. This is
apparent from the $\chi^2$ values shown in Table \ref{results} for
the same calculations fit over different angular ranges. Overall, the
results of the calculations are quite consistent. The variations obtained
when using different optical potentials were used to estimate
the uncertainties from the calculations. Weighing the calculations by
$\chi^2$ gives $C_{p_{3/2}}^{2}(^{8}Li)=0.384\pm 0.038$ fm$^{-1}$ and $%
C_{p_{1/2}}^{2}(^{8}Li)=0.048\pm 0.006$ fm$^{-1}$. Other averaging
procedures give essentially identical results. The uncertainty in
$C_{p_{3/2}}^{2}$ includes contributions from the overall normalization of
the cross section (7\%), choice of the angular range of the fit and the
optical model potentials (5\%), geometry of the neutron binding potential
used in the DWBA calculations (1.5\%), and the absolute value of the
$^{13}$C ANC
(5\%). For the smaller component, $C_{p_{1/2}}^{2}$, the
uncertainty in the fit due to different optical model potentials (8\%)
dominates.

\begin{figure}[tbp]
\includegraphics*[width=8.3cm]{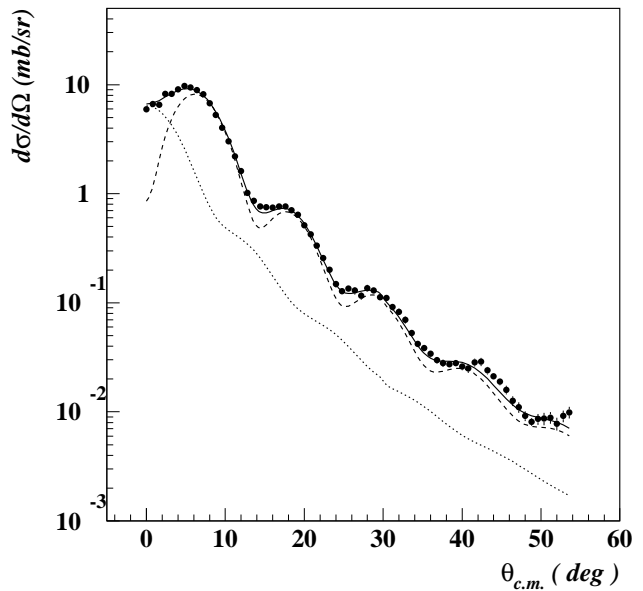}
\caption{Same as Fig.\@ \ref{8Li_angdis}, but for the first excited state in $%
^8$Li at 981 keV.}
\label{8Li_fig4}
\end{figure}

The first excited state in $^8$Li, which is the mirror of
the resonance at $E_{cm}$ = 633 keV in the $^7$Be($p$,$\gamma$)$^8$B
reaction, was also measured in the present experiment. The angular
distribution is shown in Fig.\@ \ref{8Li_fig4},
where it is compared with
a fit using the POT1 optical model parameters. The same two
components, $p_{1/2}$ $\rightarrow$ $p_{3/2}$ and $p_{1/2}$ $\rightarrow$ $%
p_{1/2}$, were calculated. The results from the fit are $%
C_{p_{3/2}}^{2}(^{8}Li^{*})=0.067\pm 0.007$ fm$^{-1}$ and $%
C_{p_{1/2}}^{2}(^{8}Li^{*})=0.015\pm 0.002$ fm$^{-1}$. The ratio of the ANCs
is $C_{p_{1/2}}^{2}/C_{p_{3/2}}^{2}(^{8}Li^{*})=0.22(3)$. Reference \cite
{grigorenko} predicts a ratio of 0.35 for this state.

To obtain the ANCs in $^{8}$B corresponding to those in $^8$Li, we use
Eq.\@ (\ref{CtoC}) and assign an additional 3\% uncertainty
to account for possible
charge-symmetry breaking effects. The ratio of the proton and neutron single
particle ANCs is $b_{p_{j}}^{2}(^{8}B)/b_{p_{j}}^{2}(^{8}Li)=1.055(20)$. This ratio was
obtained from single-particle wave functions calculated numerically for a neutron or a proton
bound in a Woods-Saxon potential with the same geometry and the same spin-orbit interaction
and with a depth adjusted to reproduce the experimental neutron or proton binding energy in
$^8$Li or $^8$B.  The potential depths were found to be nearly equal as the geometrical
parameters were varied.  
This
result is the same for both spin-orbit partners and the small uncertainty 
represents the weak dependence on the
geometry of the potential that binds the proton or neutron around its
respective core.
Inserting this ratio into Eq.\@ (\ref{CtoC}),
we find $C_{p_{3/2}}^{2}(^{8}B)=0.405\pm 0.041$
fm$^{-1}$ and $C_{p_{1/2}}^{2}(^{8}B)=0.050\pm 0.006$ fm$^{-1}$. The use of the
experimental determination of ANCs in $^{8}$Li to obtain those in $^{8}$B
was suggested in Ref.\@ \cite{natasha98} based on results of
microscopic calculations for the two nuclei, but the ratios found there are
somewhat different from the present one and their spread is considerably
larger. However, in Ref.\@ \cite{natasha98} the ratio is exaggerated because
exactly the same model wave
functions were used for the mirror nuclei ${}^{8}$B and ${}^{8}$Li. 
An evaluation within a single-particle model shows that the replacement 
of the neutron bound state wave function in the source term by the proton 
wave function leads to a decrease of the ratio by $9\%$, bringing the result 
of Ref.\@ \cite{natasha98} into agreement with the number above.

The values found for the $^{8}$B ANCs are in good agreement with those
obtained from proton transfer reactions at 12 MeV/u \cite{azhari01}, where
the average of the values extracted in two similar experiments on
different targets was found to be $C_{p_{3/2}}^{2}(^{8}B)=0.388\pm 0.039$ fm$%
^{-1}$. The two spin-orbit components could not be separated there, so
the value of 0.157 for the ratio, as predicted from a
microscopic model calculation \cite{mukh90}, was used to extract the ANCs
from the ($^{7}$Be,$^{8}$B) reactions. Changing this ratio to 0.13 decreases
the value of $S_{17}(0)$ extracted from the proton transfer reactions by
only 0.7\%.

In Ref.\@ \cite{trache01} the sum of the ANCs in $^{8}$B was extracted from
breakup reactions at intermediate energies. The value found was $%
C_{p_{3/2}}^{2}+$ $C_{p_{1/2}}^{2}=0.450\pm 0.039$ fm$^{-1}$. The present
result gives $C_{p_{3/2}}^{2}+C_{p_{1/2}}^{2}=0.455\pm 0.047$ fm$^{-1}$,
in excellent
agreement with the value from breakup. Thus the two different transfer
reactions and $^{8}$B breakup all give similar values for the astrophysical
factor, the present data giving $S_{17}(0)=17.6\pm 1.7$ eV\,b. 
This result is also in agreement, within uncertainties, with most of the
existing results for $S_{17}(0)$ from direct or indirect methods \cite
{hama,strieder,iwasa,davids}.  It is not in good agreement with the two
latest results from direct measurements \cite{jung,haas}, which
claim 
very good accuracy.  However, the present result is in good agreement
with a very recent, high precision
Coulomb dissociation study \cite{schum03} that also
calls into question the low-energy extrapolation \cite{desc94}
adopted by the recent direct measurements.
In fact, the value of $S_{17}(0)$ inferred from the measurements in
Ref.\@ \cite{haas} also agrees with our result when the extrapolation
to zero energy is done using the prescription in Ref.\@ \cite{schum03},
rather than that in Ref.\@ \cite{desc94}.

This is the first time that the rate of a direct capture reaction of
astrophysical interest has been determined through a measurement of the ANCs
in the mirror nuclear system. This represents a new variation of the
asymptotic normalization coefficient technique that will be applicable in
the future to other direct radiative transitions of astrophysical
interest
for which the proton capture ANC can be shown to
be proportional to that in the mirror system and
which
would otherwise only be accessible through experiments with
radioactive beams.

One of us (FC) acknowledges the support of the Cyclotron Institute, Texas
A\&M University, during which part of this work was completed. This work was
supported in part by the U.\thinspace S. Department of Energy under Grant
No. DE-FG03-93ER40773, the U.\,S. National Science
Foundation under Grant No.\ PHY-0140343, the Robert A. Welch 
Foundation, and EPSRC grant GR/M/82141.

\end{document}